\documentclass[]{aa}
\usepackage{times}
\usepackage{natbib}
\usepackage[dvips]{graphicx}
\usepackage{epsfig,longtable,lscape}
\usepackage{epsfig}
\usepackage{rotating}
\bibpunct[]{(}{)}{;}{a}{}{,}

\def\XMM{{\em XMM--Newton}}
\def\EPIC{{\em EPIC}}
\def\MOS{{\em MOS}}
\def\Mone{{\em MOS1}}
\def\Mtwo{{\em MOS2}}
\def\pn{{\em pn}}
\def\ROSAT{{\em ROSAT}}
\def\XTE{{\em RossiXTE}}

\def\ASM{{\em ASM}}
\def\PCA{{\em PCA}}
\def\RX{RX J1037.5--5647}

\def\TreA{{3A 0535+262}}

\begin{document}

\title{\XMM\ observation of the persistent Be/NS X--ray binary pulsar \RX\ in a low luminosity state}

\author{N. La Palombara\inst{1}, L. Sidoli\inst{1}, P. Esposito\inst{1,2}, A. Tiengo\inst{1}, S. Mereghetti\inst{1}}

\institute{INAF, Istituto di Astrofisica Spaziale e Fisica Cosmica - Milano, Via Bassini 15, I--20133, Milano, Italy
\and INFN, Sezione di Pavia, Via A.Bassi 6, I--27100, Pavia, Italy}

\titlerunning{\XMM\ observations of RX J1037.5-5647}

\authorrunning{La Palombara et al.}

\abstract{The spectra of several X--ray binary pulsars display a clear soft excess, which in most cases can be described with a blackbody model, above the main power--law component. While in the high--luminosity sources it is usually characterized by low temperature ($kT <$ 0.5 keV) and large emission radius ($R >$ 100 km), in the two persistent and low--luminosity pulsars 4U 0352+309 and RX J0146.9+6121 this component has a high temperature ($kT >$ 1 keV) and a smaller radius ($R <$ 0.5 km), consistent with the estimated size of the neutron--star polar cap. Here we report on the timing and spectral analysis of \RX, another low--luminosity persistent Be binary pulsar, based on the first \XMM\ observation of this source. We have found a best--fit period $P$ = 853.4 $\pm$ 0.2 s, that implies an average pulsar spin--up $\dot P \simeq -2\times10^{-8}$ s s$^{-1}$ in the latest decade. The estimated source luminosity is $L_{X} \sim 10^{34}$ erg s$^{-1}$, a value comparable to that of the other persistent Be binary pulsars and about one order of magnitude lower than in most of the previous measurements. The source spectrum can be described with a power law plus blackbody model, with $kT_{\rm BB} = 1.26^{+0.16}_{-0.09}$ keV and $R_{\rm BB} = 128^{+13}_{-21}$ m, suggesting a polar--cap origin of this component. These results strengthen the hypothesis that, in addition to low luminosities and long periods, this class of sources is characterized also by common spectral properties.
}

\maketitle

\section{Introduction}

Most X--ray binary pulsars (XBPs) are High Mass X--Ray Binaries (HMXRBs) in which a Neutron Star (NS) with a magnetic field B $\sim 10^{12}$ G accretes matter from a high--mass early--type star, either an OB supergiant or a Be star. The X--ray spectra of both subgroups are generally described by a rather flat power law between 0.1 and 10 keV (photon index $\sim$1) with a high--energy cutoff \citep{Nagase02}. Several of the brightest XBPs ($L_{X} \sim 10^{37-38}$ erg s$^{-1}$) have also a marked soft X--ray excess \citep{Ramsay+02,Paul+02,NaikPaul04,Yokogawa+00a,Kohno+00,Burderi+00,Manousakis+09}. A variety of simple models (both thermal and non--thermal) have been proposed to fit this feature. Only in a few cases this low--energy component unambiguously showed coherent pulses, therefore the debate over its origin remains open.

A large number of new pulsars in BeXRB systems has been discovered in the Small Magellanic Cloud, where the observations are unaffected by the high interstellar absorption along the Galactic plane \citep{Haberl&Sasaki00,Israel+00,Yokogawa+03,Macomb+03,Haberl+08,EgerHaberl08}. This allowed the detailed study of their X--ray spectra down to energies of a few hundred eV, leading to the detection of a marked soft excess in most pulsars, also in sources with relatively low luminosity ($L_{X} \sim 10^{35-36}$ erg s$^{-1}$) \citep{Yokogawa+00b,Sasaki+03,Haberl&Pietsch04,Ueno+04,Majid+04,Haberl&Pietsch05}.

Based on the these results, \citet{Hickox+04} showed that a low--energy component should be visible in all X--ray pulsars that have sufficiently high flux and small absorption. In fact, most of the soft--excess sources are at small distances and/or away from the Galactic plane (most of them are in the Magellanic Clouds). This suggests that the presence of a soft spectral component could be a very common, if not an ubiquitous, feature intrinsic to X--ray pulsars. Different explanations for its origin have been proposed, depending on the source luminosity. When $L_{X} \ge 10^{38}$ erg s$^{-1}$, the luminosity and the shape of the soft component can be explained only by the presence of an optically--thick accretion disk, which reprocesses the hard X--rays coming from the neutron star at its inner edge. In less luminous sources, with $L_{X} \le 10^{36}$ erg s$^{-1}$, the soft excess can be due to other processes, such as emission by photo--ionized or collisionally--heated diffuse gas or thermal emission from the surface of the neutron star. Finally, in the sources of intermediate luminosity, either or both of these types of emission can be present.

Recently, a clear thermal excess has been observed also in the cases of RX J0146.9+6121 \citep{LaPalombaraMereghetti2006} and 4U 0352+309 \citep{LaPalombaraMereghetti2007}, which are two persistent Be pulsars. This is a specific class of binary pulsars characterized by low luminosity ($L_{X} \sim 10^{34-35}$ erg s$^{-1}$) and long pulse period ($P >$ 100 s); moreover, they show no transient behavior, since the source luminosity varies at most of a factor $\sim$ 10 \citep{Negueruela98}. These properties suggest that the Be star orbits the NS in a wide and nearly circular orbit, continuously accreting material from the low--density outer regions of the circumstellar envelope. This picture has been confirmed for 4U 0352+309 by the discovery of an orbital period of 250.3 days \citep{Delgado-Marti+2001}. The detection of the thermal excess in these two persistent pulsars was favoured by their small distance ($d \le$ 2.5 kpc) and interstellar absorption ($N_{\rm H} \sim 10^{21}$ cm$^{-2}$).

It is interesting to investigate if other persistent Be binary pulsars are characterized by the same type of spectral feature, in order to check if it is a common property of this class of sources. To this aim, we have observed with \XMM\ the Be/NS binary pulsar \RX, identified with LS 1698, a B0 III--Ve star at $\sim$ 5 kpc \citep{Motch+97}. This source is likely associated to the \textit{UHURU} source 4U 1036-56 \citep{Forman+78}, the \textit{ARIEL V} source 3A 1036-565 \citep{Warwick+81} and the \textit{OSO 7} source 1M 1022-554 \citep{Markert+79}. Thanks to observations performed by the \textit{PCA} instrument on--board \XTE, \citet{ReigRoche99} performed the first detailed timing and spectral analysis, and discovered a pulsation with period $P = 860 \pm 2$ s. The source spectrum could be described by a cut--off power law, with a low cut--off energy ($E_{\rm cut} = 4.7$ keV) and a weak (equivalent width EQW $\simeq$ 65 eV) iron line at 6.5 keV; moreover, the estimated source luminosity ($L_{\rm X} \simeq 2\times 10^{35}$ erg s$^{-1}$) was fully consistent with the luminosity level detected by the previous observations, and the light curve showed no large variability. Therefore, \RX\ is a long--period, low--luminosity Be/NS XBP which has been detected every time it has been observed. All these properties are typical of the persistent Be/NS pulsars, therefore \citet{ReigRoche99} suggested that \RX\ is a potential member of this class of sources.

\section{Observations and data reduction}\label{sec:2}

\RX~was observed by \XMM\ on 2009 January 9. The three \EPIC~cameras, i.e. one \pn\ \citep{Struder+01} and two \MOS\ \citep{Turner+01}, were operated in \textit{Small Window} mode, with a time resolution of 6 ms for the \pn\ camera and of 0.3 s for the two \MOS\ cameras; the medium thickness filter was used. The effective source exposure time was $\sim$ 28 ks for the two \MOS\ and $\sim$ 20 ks for the \pn, since the use of the \textit{Small Window} mode for this camera implies a $\sim$ 30 \% dead time. Also the \textit{Reflection Grating Spectrometer} was used but, due to the source faintness, the count statistics was very low; therefore we ignored its spectra.

We used  version 8.0 of the \XMM~{\em Science Analysis System} (\texttt{SAS}) to process the event files.  After the standard pipeline processing, we looked for possible periods of high instrumental background, caused by flares of soft protons with energies less than a few hundred keV. We found that the last $\sim$ 8 ks of the observation were affected by a high soft--proton contamination, therefore we rejected the data of this part of the observation in the spectral analysis. The resulting effective exposure time was 14.5 and 20 ks for the \pn\ and the \MOS\ cameras, respectively.

\section{Timing analysis}\label{timing}

We selected source counts within a circular region around the source position, with extraction radii of 20 and 15$''$ for the \pn\ and the two \MOS\ cameras, respectively. Since in \textit{Small Window} mode a very small sky area is imaged by the instrument, we were forced to select the corresponding background data on small circular areas, with radii of 20$''$. We accumulated all the events with pattern range 0--4 (i.e. mono-- and bi--pixel events) and 0--12 (i.e. from 1 to 4 pixel events) for the \pn~and the two \MOS~cameras, respectively. This selection resulted in a net \textit{count rate} (CR) of $\sim$ 0.3 cts s$^{-1}$ for the \pn\ camera and $\sim$ 0.13 cts s$^{-1}$ for each of the two \MOS\ cameras.

To measure the pulse period, we converted the event arrival times to the solar system barycenter, for each instrument separately; then, in order to increase the count statistic, we combined the three datasets and measured the pulse period by a standard phase--fitting technique \citep{dallosso+03}, obtaining a best--fit period \textit{P} = 853.4 $\pm$ 0.2 s.

\begin{figure}[h]
\centering
\resizebox{\hsize}{!}{\includegraphics[angle=-90,clip=true]{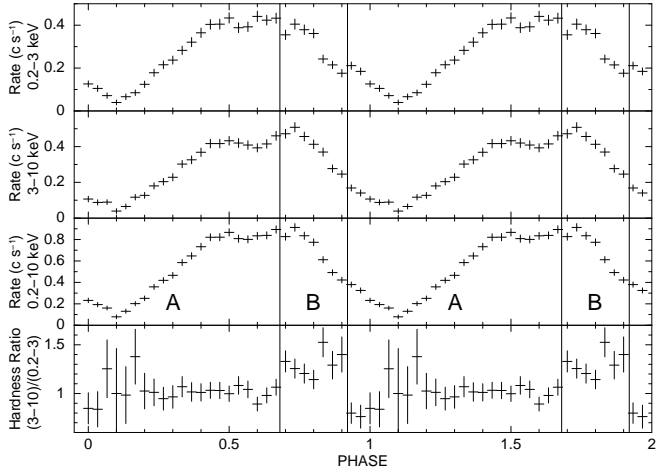}}
\caption{Background--subtracted light curves of \RX~in the energy ranges 0.2--3, 3--10, and 0.2--10 keV, folded at the best--fit period \textit{P} = 853.4 s. The vertical lines indicate the phase intervals used for the phase--resolved spectroscopy (see Section 5).}
\label{flc}
\end{figure}

Thanks to the \XMM\ observation, we could investigate the pulse profile, for the first time, even at energies below 3 keV, which could not be studied by \XTE. In Fig.~\ref{flc} we show the folded light curves in the three energy intervals, 0.2--3 (soft), 3--10 (hard), and 0.2--10 keV, together with the folded \textit{hardness--ratio} HR between the hard (H) and soft (S) light curves (computed as H/S). The shape of the pulse profile is similar in the three energy ranges. The measured pulsed fraction in the 0.2--10 keV energy band, defined as (CR$_{\rm max}$ - CR$_{\rm min}$)/(2 $\times$ CR$_{\rm average}$), is 75 \%. The HR is rather constant along most of the pulse period; it increases only at the end of the pulse maximum, when the source count rate starts to decrease.

\begin{figure*}[t]
\centering
\resizebox{\hsize}{!}{\includegraphics[angle=-90,clip=true]{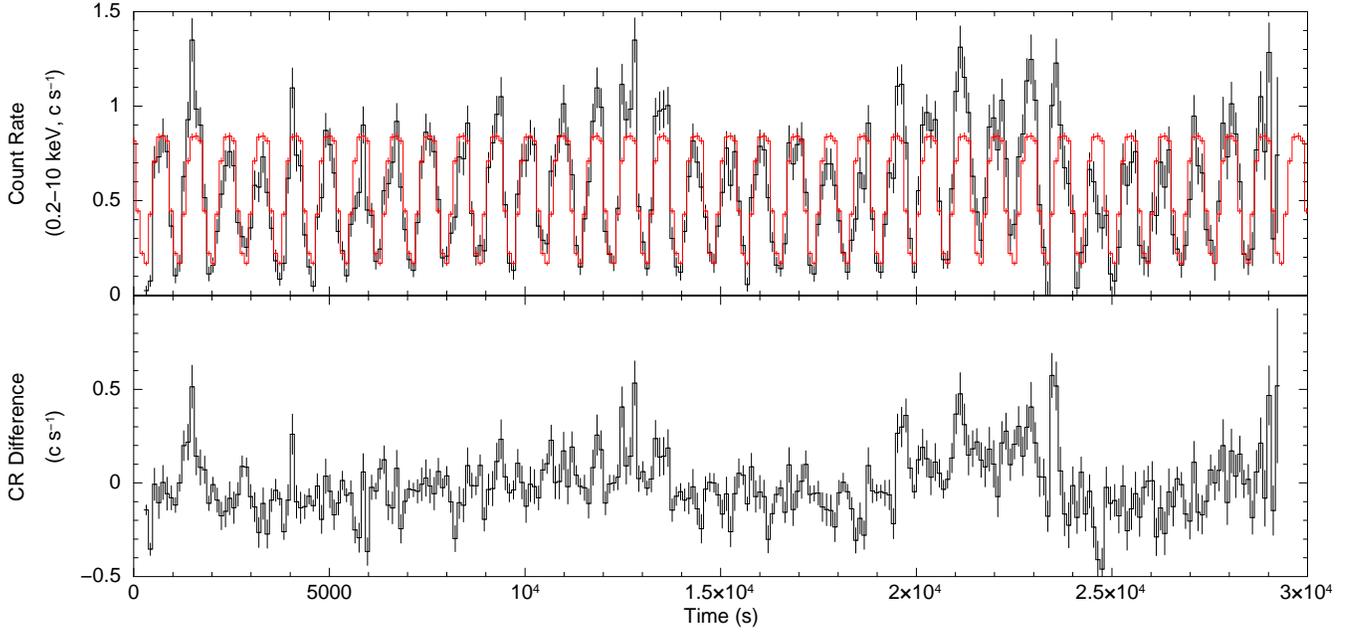}}
\caption{\textit{Top}: observed (\textit{black}) and average (\textit{red}) light curve of \RX\ in the energy range 0.2--10 keV; \textit{Bottom}: count--rate difference between the individual time bins of the observed and average light curve.}
\label{lc}
\end{figure*}

\begin{figure}[h]
\centering
\resizebox{\hsize}{!}{\includegraphics[angle=-90,clip=true]{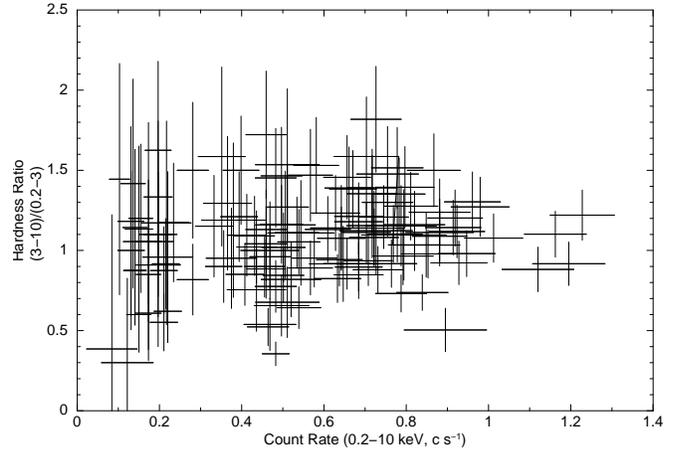}}
\caption{Hardness--ratio variation of \RX~as a function of the 0.2--10 keV count rate, with a time bin of 213 s.}
\label{hr_cr_sum}
\end{figure}

It is interesting to compare the light curve of the whole \XMM\ observation with the \textit{average} one, which is obtained assuming that all the pulses are equal to the average folded light curve. To this aim, we considered a binning time of 106.5 s, corresponding to eight bins for each pulse period. In Fig.~\ref{lc} we show the total (\pn+\Mone+\Mtwo) background--subtracted light curve, in the 0.2--10 keV energy range, and the average one, together with their difference. It shows that there are significant differences between the individual pulses, as well as some variability on longer timescales. Moreover, we checked that the source HR between the hard and soft bands is rather constant around the average value of 1.08 ($\chi^{2}_{\nu}$/d.o.f. = 1.89/134), and does not depend on the source count rate (Fig.~\ref{hr_cr_sum}).

\section{Spectral analysis}\label{sec:3}
For the source and background spectra, we considered the same extraction parameters we used for the light curves. We generated the applicable response matrices and ancillary files using the \texttt{SAS} tasks \texttt{rmfgen} and \texttt{arfgen}. In order to ensure the applicability of the $\chi^{2}$ statistics all spectra were rebinned with a minimum of 30 counts per bin; they were fitted in the energy range 0.2--10 keV using \texttt{XSPEC} 11.3.2. In the following, all spectral uncertainties and upper--limits are given at 90 \% confidence level for one interesting parameter.

After checking that separate fits of the three cameras gave consistent results, we fitted them simultaneously, in order to increase the statistics; to this aim, we introduced relative normalization factors among the spectra of the three cameras. Using an absorbed power--law (\textit{PL}) model, we obtained a hydrogen column density $N_{\rm H} = (6.2\pm0.5)\times 10^{21}$ cm$^{-2}$ and a photon--index $\Gamma$ = 0.94$\pm$0.03, with $\chi^{2}_{\nu}$/d.o.f. = 1.30/277. On the other hand, using an absorbed blackbody (\textit{BB}) model, we obtained a hydrogen column density $N_{\rm H} = (0.9\pm0.2)\times 10^{21}$ cm$^{-2}$ and a \textit{BB} temperature $kT_{\rm BB} = 1.52 \pm 0.03$ keV, with $\chi^{2}_{\nu}$/d.o.f. = 1.20/277; assuming a source distance of 5 kpc we obtained a \textit{BB} radius $R_{\rm BB} = 131 \pm 4$ m.

\begin{figure}[h]
\centering
\resizebox{\hsize}{!}{\includegraphics[angle=-90,clip=true]{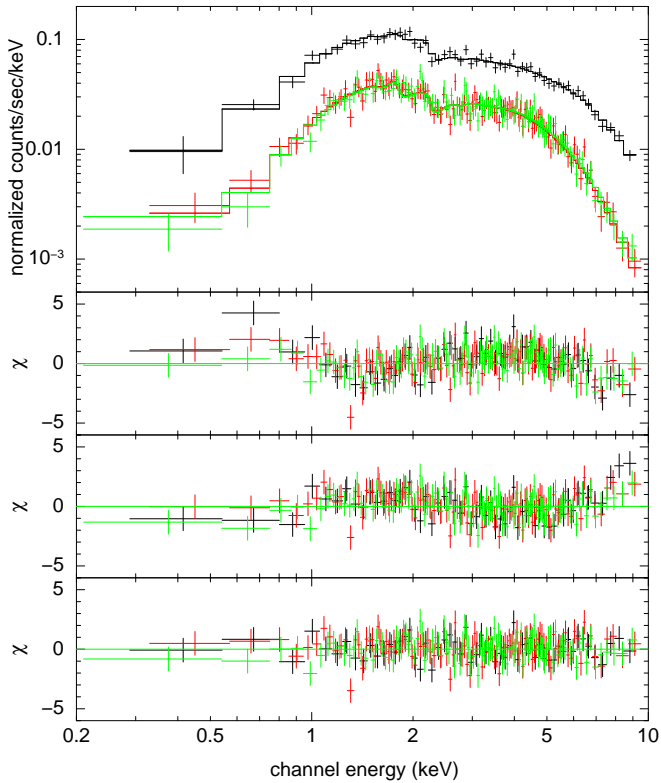}}
\caption{\textit{Top panel}: total spectrum of \RX~with the best--fit \textit{PL+BB} model. The spectra of the \pn, \Mone\ and \Mtwo\ cameras are shown in black, red and green, respectively. \textit{Middle panels}: residuals (in units of $\sigma$) between data and model in the case of the single \textit{PL} and of the single \textit{BB}. \textit{Bottom panel}: residuals in the case of the \textit{PL+BB}.}\label{all_spetrum_powbb}
\end{figure}

In both cases the fit of the spectrum with a single--component model shows large residuals, therefore we repeated the fit with a \textit{PL+BB} model. In this way we obtained a significant improvement of the fit quality (Fig.~\ref{all_spetrum_powbb}), since we obtained $\chi^{2}_{\nu}$/d.o.f = 0.88/275; the corresponding best--fit parameters are $N_{\rm H} = (2.8\pm0.9) \times 10^{21}$ cm$^{-2}$, $\Gamma = 0.51^{+0.17}_{-0.29}$ and $kT_{\rm BB} = 1.26^{+0.16}_{-0.09}$ keV. In comparison with the single \textit{PL} and \textit{BB} model, the F--test analysis provided a probability, respectively, \textit{P} = 1.26 $\times$ 10$^{-24}$ and \textit{P} = 9.95 $\times$ 10$^{-20}$ that the improvement of the fit occurs by
chance. Both components are significant at 99 \% confidence level: the \textit{PL} normalization is $I_{\rm PL} = 7.1^{+4.5}_{-4.6}\times10^{-5}$ ph cm$^{-2}$ s$^{-1}$ keV$^{-1}$ at 1 keV and, assuming a source distance of 5 kpc, we obtained a radius $R_{\rm BB} = 128^{+13}_{-21}$ m for the \textit{BB} component. The unabsorbed flux in the energy range 0.2--10 keV is $f_{\rm X}\sim 4\times10^{-12}$ erg cm$^{-2}$ s$^{-1}$, about 42 \% of which is due to the \textit{BB} component; this translates into a source luminosity $L_{\rm X}\simeq 1.2\times10^{34}$ erg s$^{-1}$.

We looked also for narrow iron K$_{\alpha}$ emission lines between 6 and 7 keV, with different widths between 0 and 0.5 keV. We found no evidence for such a component, with an upper limit on its equivalent width of 0.2 keV (at 90 \% c.l.) at most.

\section{Phase--resolved spectroscopy}\label{spectroscopy}

In the source folded light curve (Fig.~\ref{flc}) we observed that the source spectrum becomes slightly harder at the end of the pulse maximum. In order to study the source behavior in more detail, we analyzed the background subtracted spectra in two different phase intervals, i.e. $\phi$ = 0--0.68, 0.92--1 (phase interval A) and $\phi$ = 0.68--0.92 (phase interval B) 

\begin{figure}[h]
\centering
\resizebox{\hsize}{!}{\includegraphics[angle=-90,clip=true]{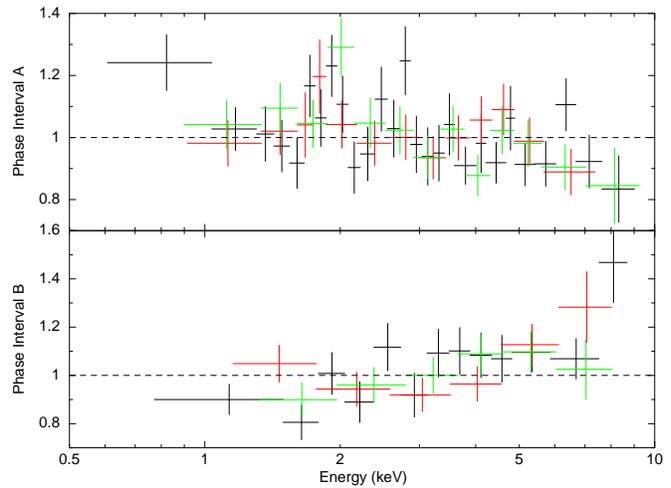}}
\caption{Ratios of the spectra corresponding to the phase intervals shown in Fig.~\ref{flc} to the average model spectrum. The parameters are fixed at the best--fit values for the total spectrum, except for the normalizations, which are at their best--fit values for each phase interval.}\label{ratios}
\end{figure}

We fit the two background--subtracted spectra with the best--fit \textit{PL+BB} model of the phase--averaged spectrum, leaving only the relative normalization factors between the two phases free to vary; in Fig.~\ref{ratios} we report the ratios of the two spectra of each instrument to this renormalized average model. The data of the three detectors show that the spectra of phase intervals A and B are, respectively, softer and harder than the phase--averaged one, which implies a spectral variability along the pulse phase.

We then fitted the two spectra independently, using both a single \textit{PL} model, a single \textit{BB} model and a \textit{PL+BB} model; their best--fit parameters are reported in Table~\ref{2spectra_fit}. In both cases we found that, as for the phase--averaged spectrum, the \textit{BB} model provides a better fit than the \textit{PL} model and that the use of a \textit{PL+BB} model improves significantly the spectral fit goodness, since there is a large decrease in the $\chi^{2}$ value and the F--test analysis gives a negligible probability that the improvement of the fit occurs by chance. Irrespective of the adopted model, we always found that the absorbing hydrogen column density is consistent with a constant value, as expected if it is due to the interstellar extinction. With both the single--component and the two--component models the photon index shows evidence for a variability between the two phase intervals, thus confirming the spectral variability; on the other hand, the \textit{BB} temperature and radius are variable only in the case of the single--component model, while they are rather constant in the case of the two--component model, since there is a full agreement for the emission radii and also the temperatures are comparable.

\begin{table*}[htbp]
\caption{Best--fit spectral parameters for the phase--resolved spectroscopy of \RX, in the case of the independent fit of the two spectra. $N_{\rm H}$, $kT_{\rm BB}$, and $R_{\rm BB}$ are measured in units of $10^{21}$ cm$^{-2}$, keV, and meters (for a source distance of 5 kpc), respectively. $P_{F-test}(PL)$ and $P_{F-test}(BB)$ are the probability that the improvement of the fit, compared to the single \textit{PL} and \textit{BB} model, respectively, occurs by chance.}\label{2spectra_fit}
\begin{center}
\begin{tabular}{c|cc|cc|cc} \hline
Phase interval		&	A		&	B		&	A		&	B		&	A		&	B		\\ \hline
$N_{\rm H}$		& 6.6 $\pm$ 0.7		& 6.6$^{+1.5}_{-1.3}$	& 0.5 $\pm$ 0.3		& 0$^{+0.8}_{-0.0}$	& 2.1$^{+1.4}_{-1.1}$	& 0.9$^{+0.9}_{-0.8}$	\\ \hline
$\Gamma$		& 1.01 $\pm$ 0.06	& 0.79 $\pm$ 0.10 	&			&			& 0.4$^{+0.5}_{-1.0}$	& -2.5$^{+0.3}_{-0.2}$	\\
$f_{\rm PL}^{a}$	& 4.8 (100 \%)		& 5.5 (100 \%)		&			&			& 2.0 (51 \%)		& 1.5 (31 \%)		\\ \hline
$kT_{\rm BB}$		&			&			& 1.45 $\pm$ 0.03	& 1.69$^{+0.06}_{-0.08}$& 1.22$^{+0.18}_{-0.09}$& 1.44$^{+0.12}_{-0.22}$\\
$R_{\rm BB}$		& 			&			& 139 $\pm$ 6		& 117$^{+9}_{-6}$	& 142$^{+29}_{-24}$	& 141$^{+11}_{-13}$	\\
$f_{\rm BB}^{a}$	&			&			& 3.4 (100 \%)		& 4.1 (100 \%)		& 1.9 (49 \%)		& 3.4 (69 \%)		\\ \hline
$f_{\rm TOT}^{a}$	& 4.8			& 5.5			& 3.4			& 4.1			& 3.9			& 4.9			\\ 
$\chi^{2}_{\nu}$/d.o.f.	& 2.00/80		& 1.50/33		& 1.45/80		& 1.17/33		& 1.05/78		& 0.80/31		\\
$P_{F-test}(PL)$	&			&			&			&			& 4.79$\times 10^{-12}$	& 2.15$\times 10^{-5}$	\\
$P_{F-test}(BB)$	&			&			&			&			& 1.33$\times 10^{-6}$	& 1.10$\times 10^{-3}$	\\ \hline
\end{tabular}
\end{center}
\begin{small}
$^{a}$ Unabsorbed flux in the energy range 0.2--10 keV, in units of $10^{-12}$ erg cm$^{-2}$ s$^{-1}$
\end{small}
\end{table*}


To investigate the relative variations of the two components with the period phase, we also simultaneously fitted the two spectra forcing the same value of $N_{\rm H}$, $\Gamma$ and $kT_{\rm BB}$ for the two phase intervals. In this case we obtained $N_{\rm H} = 1.4^{+0.6}_{-0.5}\times10^{21}$ cm$^{-2}$, $\Gamma_{\rm PL}=-0.2^{+0.3}_{-0.9}$ and $kT_{\rm BB} = 1.21^{+0.07}_{-0.06}$ keV, with $\chi^{2}_{\nu}$/d.o.f. = 0.99/115; the corresponding normalization values are reported in Table~\ref{2spectra_common}. With these constraints on the model parameters, the spectral changes as a function of the phase are reproduced by the variations in the relative contribution of the two components. The values reported in Table~\ref{2spectra_common} show that almost all of the spectral variability is due to the \textit{PL} component, since its flux changes of a factor $\sim$ 2 between the two phase intervals; on the other hand, there is only a 15 \% variation of the \textit{BB} flux, since we obtain comparable values for the emission radius. However, this finding is affected by large errors on the normalization of the \textit{PL} component.

As further check, we modified the test model by linking together the \textit{BB} parameters for the two spectra, while both the photon--index $\Gamma$ and the normalization $I_{\rm PL}$ of the \textit{PL} component could vary independently in the two phase intervals. The resulting best fit gives $N_{\rm H}= (2.3\pm1.0)\times10^{21}$ cm$^{-2}$, $kT_{\rm BB} = 1.25^{+0.13}_{-0.09}$ keV and $R_{\rm BB}$ = $137^{+13}_{-21}$ m, while the power--law parameters are shown in Table~\ref{2spectra_common}. Even with this model we found a very good fit quality ($\chi^{2}_{\nu}$/d.o.f. = 0.96/115), fully comparable to that obtained in the previous case. This result suggests that it is possible to attribute the whole spectral variability to the \textit{PL} component. However, we note that the power--law parameters are not well constrained, since they are always affected by large errors. Moreover, for the spectrum of the phase interval B the independent fit with a \textit{PL+BB} model provides a large negative value of $\Gamma$, which is very different to that obtained from any other fit. This result can be explained by the large contribution of the \textit{BB} component to the total flux, while in the other cases most of the flux is due to the \textit{PL} component: in fact, for the phase interval B we can obtain a good fit also using a simple \textit{BB} model ($\chi^{2}_{\nu}$/d.o.f. = 1.17/33). This means that an additional \textit{PL} component is required only to fit the high energy end of the spectrum, thus providing a large negative value of the photon index.

\begin{table}[htbp]
\caption{Best--fit values for the black--body and power--law parameters, when the two spectra are fitted simultaneously with common values of $N_{\rm H}$, $\Gamma$ and $kT_{\rm BB}$ (\textit{case 1}) or with common values of $N_{\rm H}$, $kT_{\rm BB}$ and $R_{\rm BB}$ (\textit{case 2}).}\label{2spectra_common}
\begin{tabular}{c|cc|cc} \hline
Phase	 		&		\multicolumn{2}{c}{case 1}					&		\multicolumn{2}{c}{case 2}	\\
interval 		&		A 			&		B			&	A				&	B				\\ \hline
$N_{\rm H}^{a}$		& \textit{1.4$^{+0.6}_{-0.5}$}		& \textit{1.4$^{+0.6}_{-0.5}$}		& \textit{2.3 $\pm$ 1.0}		& \textit{2.3 $\pm$ 1.0}		\\ \hline
$\Gamma$		& \textit{-0.2$^{+0.3}_{-0.9}$}		& \textit{-0.2$^{+0.3}_{-0.9}$}		& 0.51$^{+0.35}_{-0.57}$		& 0.09$^{+0.37}_{-0.66}$		\\
$I_{\rm PL}^{b}$	& 1.2$^{+2.0}_{-1.0}$			& 2.4$^{+3.2}_{-2.0}$			& 6.0$^{+5.1}_{-4.8}$			& 4.3$^{+4.1}_{-3.2}$			\\
$f_{\rm PL}^{c}$	& 1.5 (39 \%)				& 2.9 (60 \%)				& 2.0 (51 \%)				& 2.9 (61 \%)				\\ \hline
$kT_{\rm BB}$ (keV)	& \textit{1.21$^{+0.07}_{-0.06}$}	& \textit{1.21$^{+0.07}_{-0.06}$}	& \textit{1.25$^{+0.13}_{-0.09}$}	& \textit{1.25$^{+0.13}_{-0.09}$}	\\
$R_{\rm BB}^{d}$	& 160$^{+11}_{-8}$			& 147$^{+16}_{-19}$			& \textit{137$^{+13}_{-21}$}		& \textit{137$^{+13}_{-21}$}		\\
$f_{\rm BB}^{c}$	& 2.3 (61 \%)				& 2.0 (40 \%)				& 1.9 (49 \%)				& 1.9 (39 \%)				\\ \hline
$f_{\rm TOT}^{c}$	& 3.8					& 4.9					& 3.9					& 4.8					\\
$\chi^{2}_{\nu}$/d.o.f.	& 	\multicolumn{2}{c}{0.99/115}						& 	\multicolumn{2}{c}{0.96/115}						\\ \hline
\end{tabular}
\begin{small}
\\
$^{a}$ Hydrogen column density in units of $10^{21}$ cm$^{-2}$

$^{b}$ Intensity of the power--law component in units of $10^{-5}$ ph cm$^{-2}$ s$^{-1}$ keV$^{-1}$ at 1 keV

$^{c}$ Unabsorbed flux in the energy range 0.2--10 keV, in units of $10^{-12}$ erg cm$^{-2}$ s$^{-1}$

$^{d}$ Radius of the black--body component (in metres) for a source distance of 5 kpc.
\end{small}
\end{table}

\section{Discussion}\label{sec:6}

Before the \XMM\ observation of \RX\, the only X--ray investigation of this source was performed by \XTE\ \citep{ReigRoche99}. Therefore it is interesting to compare our results with those obtained ten years before.

From the timing analysis, we have obtained a new refined pulse period, \textit{P} = 853.4 $\pm$ 0.2 s, which is lower than the previous value of 860 $\pm$ 2 s found by \XTE. This implies a pulsar spin--up during the past ten years, with $\dot P = (-1.91\pm0.58)\times10^{-8}$ s s$^{-1}$, and suggests a momentum transfer to the neutron star.

Based on the observed source flux, we have estimated a source luminosity $L_{\rm X} \sim 10^{34}$ erg s$^{-1}$ in the 2--10 keV energy range (assuming a source distance of 5 kpc), which is one order of magnitude lower than the luminosity level detected by \XTE. From the spectral analysis, we have obtained a hydrogen column density $N_{\rm H}$ = (2.8 $\pm$ 0.9)$\times10^{21}$ cm$^{-2}$, which is much lower than the values estimated by \XTE: the lowest result (obtained with a double BB model) was (2.1 $\pm$ 0.3)$\times10^{22}$ cm$^{-2}$, which is almost one order of magnitude higher than our result; moreover, using our same \textit{PL+BB} model, the \XTE\ data provided a best--fit value of (8.2 $\pm$ 0.5)$\times10^{22}$ cm$^{-2}$, which is $\sim$ 30 times higher than our value. On this subject, we emphasize that the energy range of the \XTE\ spectral analysis (above 3 keV) is not well suited for a good estimate of $N_{\rm H}$, while the low--energy end of the \XMM\ spectra (0.2 keV) allows a much more reliable analysis. Moreover, we also note that LS 1698, the optical counterpart of \RX, has a color excess \textit{E(B-V)} $\simeq$ 0.75 \citep{Motch+97}; assuming $A_{\rm V}$ = 3.1 \textit{E(B-V)} and the average relation $A_{\rm V}$ = $N_{\rm H} \times 5.59 \times 10^{-22}$ cm$^{-2}$ between optical extinction and X--ray absorption \citep{PredehlSchmitt95}, this would predict $N_{\rm H}=4.16\times10^{21}$ cm$^{-2}$, a value comparable to our result. This indicates that in the \XMM\ observation the measured absorption is due to the interstellar medium and not to local matter around the system, which has a negligible effect, while in the \XTE\ observations also intrinsic absorption is possible.

The \XTE\ spectrum analyzed by \citet{ReigRoche99} was fitted with the same model (\textit{PL+BB}), resulting in a quite steeper power--law component ($\Gamma$ = 1.84 $\pm$ 0.06 instead of $0.51^{+0.17}_{-0.29}$) and in a blackbody component with a much higher temperature ($kT$ = 2.9 $\pm$ 0.2 keV instead of $kT$ = 1.26$^{+0.16}_{-0.09}$ keV). Even in this case, it is possible that these differences arise not only because the spectral analysis was performed in two different energy ranges, but also because the source was one order of magnitude brighter during the \XTE\ observation. On the other hand, the two analysis provided comparable values for the emission radius of the blackbody component (100 $\pm$ 10 m for \XTE\ and $128^{+13}_{-21}$ m for \XMM).

We found no evidence of an Fe--K line in the source spectrum, with an upper limit of 0.2 keV to its equivalent width (at 90 \% c.l.), while \citet{ReigRoche99} report a gaussian line at E = 6.5 $\pm$ 0.2 keV, with an equivalent width $\simeq$ 65 eV. It is very likely that \XMM\ missed a clear detection of this feature due to the different luminosity level of the source, since a lower accretion rate on the NS implies a smaller amount of material around it and, then, a reduced probability for the emerging X--rays to produce an iron line feature.

The comparison of the pulse profiles in two energy bands suggests a possible spectral variability of the source with the pulse phase, therefore we performed a phase--resolved spectral analysis, which previously had never been carried out for this source. The fit of two phase--resolved spectra with the best--fit model of the phase--averaged one confirmed the suspected variability. The independent fit of these spectra with different emission models showed that the absorbing hydrogen column density is nearly constant along the pulse phase, as expected if it is due to the interstellar extinction; therefore the observed variability must be due to the intrinsic source emission. The same fit proved also that, for both spectra, both the thermal and non--thermal component are necessary in order to well reproduce the observed data. Finally, the simultaneous fit of the two spectra with a common value of $N_{\rm H}$ shows that the whole spectral variability can be attributed to the \textit{PL} component, since the \textit{BB} one is nearly constant along the pulse.

\citet{ReigRoche99} suggested that \RX\ is a persistent Be binary pulsars, since this source has been detected every time it has been observed and does not show large outbursts. In order to investigate the long--term evolution of \RX, we considered all the X--ray flux measurements obtained for this source since its discovery in the early '70s. For the \textit{UHURU} \citep{Forman+78}, \textit{OSO 7} \citep{Markert+79} and \textit{ARIEL V} \citep{Warwick+81} detections, we used the published CR values and the appropriate count--rate--to-flux conversion factors to obtain the corresponding 2--10 keV fluxes. In the case of the \XTE\ \textit{All Sky Monitor} (\ASM), we considered the day--by--day light curve, that covers the time interval from January 1996 to March 2009. Since the detection sensitivity of this instrument is very low, in several cases it failed to detect \RX, owing to its low flux: for this reason, in order to estimate the average flux level of the source, we binned the data in five different time bins of 1000 days each; we calculated the average CR in each bin and, assuming our best--fit \textit{PL+BB} emission model, we used the \textit{HEASARC} tool \textit{WebPIMMS} to obtain the corresponding average flux between 2 and 10 keV. On the other hand, for the data of the \XTE\ \textit{Photon Counting Array} (\PCA), we based on the results obtained by \citet{ReigRoche99} to calculate the source flux in the same energy range. Finally, we also considered the CRs measured by \ROSAT, which were obtained during the \textit{All Sky Survey} \citep{Voges+99} and in a following, specific observation \citep{Motch+97}. In their case, we had to infer the source flux in the 2--10 keV energy band from the CR measured in the 0.1--2.4 keV range: to this aim, we used again the \textit{WebPIMMS} tool and assumed our best--fit \textit{PL+BB} emission model.

\begin{figure}[h]
\centering
\resizebox{\hsize}{!}{\includegraphics[angle=-90,clip=true]{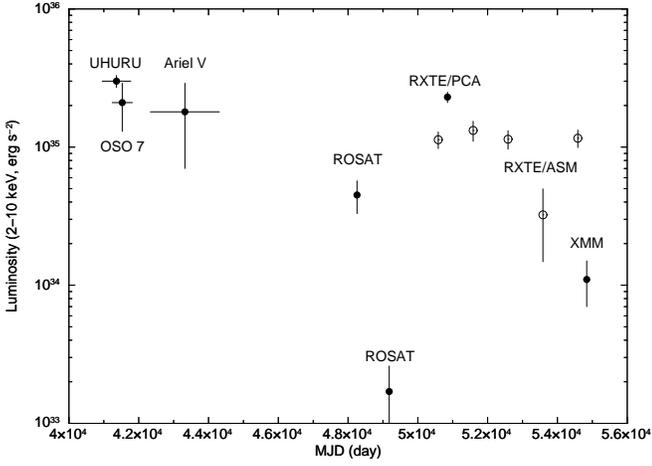}}
\caption{Luminosity history of \RX\ from 1971 to 2009. Reported luminosities are based on the unabsorbed flux in the 2--10 keV and on a source distance of 5 kpc \citep{Motch+97}. Open circles refer to the average luminosity (between 2 and 10 keV) estimated by \XTE\ \textit{ASM} in five time intervals of 1000 days each.}\label{luminosity_evolution}
\end{figure}

Assuming a source distance of 5 kpc \citep{Motch+97}, we used the calculated fluxes to estimate the corresponding luminosities, which are reported in Fig.~\ref{luminosity_evolution}. It shows that, in most observations performed before \XMM, \RX\ was detected at a luminosity level $L_{\rm X} = (1-3) \times 10^{35}$ erg s$^{-1}$ between 2 and 10 keV; moreover, also the average luminosity observed by the \textit{ASM} is $\sim 10^{35}$ erg s$^{-1}$. This is a remarkable behavior, since the performed observations span a time interval of about 35 years, and confirms that \RX\ is a persistent, low--luminosity binary pulsar. However, the observation performed by \XMM\ detected the source at a luminosity level $L_{\rm X} \sim 10^{34}$ erg s$^{-1}$, which is one order of magnitude lower than the previous average value. Moreover, in one of the \ROSAT\ observations it was detected with an even lower luminosity. It is possible that the \ROSAT\ results are due to an inaccurate estimate of the source flux, since it is obtained from a CR measurement in a different energy band, but even such an error could hardly explain the very low luminosity value obtained in the second \ROSAT\ observation. These results confirm that \RX\ is a persistent source, since it has been detected every time it has been observed, but show also that the source is characterized by a significant variability.


\begin{figure}[h]
\centering
\resizebox{\hsize}{!}{\includegraphics[angle=-90,clip=true]{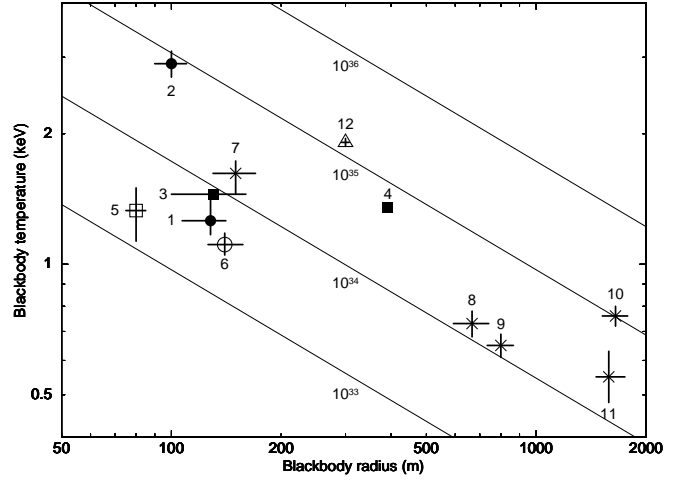}}
\caption{Best--fit values for radius and temperature of the \textit{BB} component in the case of \RX\ (\textit{filled circles}), 4U 0352+309 (\textit{filled squares}), RX J0146.9+6121 (\textit{open circle}), \TreA\ (\textit{open square}), SAX J2103.5+4545 (\textit{open triangle}), and 4U 2206+54 (\textit{crosses}). The continuous lines connect the blackbody parameters corresponding to four different levels of luminosity (in erg s$^{-1}$). References: 1 - this work; 2 - \citet{ReigRoche99}; 3 - \citet{Coburn+01}; 4 - \citet{LaPalombaraMereghetti2007}; 5 - \citet{MukherjeePaul05}; 6 - \citet{LaPalombaraMereghetti2006}; 7 - \citet{Masetti+04}; 8,9,10 - \citet{Torrejon+04}; 11 - \citet{Reig+09}; 12 - \citet{Inam+04}.}\label{BBparameters}
\end{figure}

A \textit{BB} excess above the main power--law component has been detected also in the case of LS I +61$^{\circ}$ 235/RX J0146.9+6121 \citep{LaPalombaraMereghetti2006} and X Persei/4U 0352+309 \citep{Coburn+01,LaPalombaraMereghetti2007}, which are the two main members of the class of the persistent Be/NS XBPs. A similar feature was observed also during the low--luminousity quiescent state of the transient Be/NS XBP \TreA\ \citep{Orlandini+04,MukherjeePaul05}, which is comparable to the previous persistent Be pulsars for its low luminosity ($L_{\rm X}\le10^{35}$ erg s$^{-1}$) and long pulse period ($P$ = 103 s); moreover, a blackbody component has been observed also in 4U 2206+54 \citep{Masetti+04,Torrejon+04,Reig+09}, a high--mass X--ray binary with a comparable luminosity and a very long pulse period ($P$ = 5560 s), and in the X--ray pulsar ($P$ = 358.6 s) SAX J2103.5+4545 \citep{Inam+04}. In Fig.~\ref{BBparameters} we report the best--fit radius and temperature for these sources, together with lines showing four different levels of the blackbody luminosity. In the case of \RX\ (\textit{filled circles}) and 4U 0352+309 (\textit{filled squares}), we report two set of values, since both sources were observed by \XTE\ and \XMM\ at two different luminosity levels ($L_{\rm X} \sim 10^{34}$ and $\sim 10^{35}$ erg s$^{-1}$, respectively): in the case of \RX\ the low and high luminosity were observed by \XMM\ and \XTE\, respectively, while for 4U 0352+309 it was the opposite. We report various measurements also for 4U 2206+54 (\textit{crosses}), corresponding to different observations. The figure shows that the results obtained for \RX\ are in full agreement with those obtained by various observations of the previous sources, since their spectral parameters are clustered in a narrow range of values, i.e. $kT_{\rm BB} \sim$ 1--1.5 keV and $R_{\rm BB} \sim$ 100 m. We emphasize that, in all these cases, the estimated total source X--ray luminosity is $\sim 10^{34}$ erg s$^{-1}$, with a 20--40 \% contribution of the blackbody component. For the persistent Be pulsars \RX\ and 4U 0352+309, spectral values outside the previous range (points 2 and 4) were obtained by the \XMM\ observation of 4U 0352+309 and the \XTE\ observation of \RX, when the sources were detected at a higher luminosity level ($L_{\rm X} \sim 10^{35}$ erg s$^{-1}$); however, we note that comparable blackbody parameters were obtained also for SAX J2103.5+4545, which is characterized by a comparable luminosity. On the other hand, in most cases the non--Be binary pulsar 4U 2206+54 shows a different behavior also when observed at a comparable luminosity level: $kT_{\rm BB} <$ 1 keV and $R_{\rm BB} >$ 500 m (points 8 to 11). Finally, we note that the \XMM\ spectrum of \RX\ is characterized by the hardest \textit{PL} component among the previous sources, since it has a photon--index $\Gamma$ = 0.5, while $\Gamma \ge 1$ in all the other sources.

\begin{figure}[h]
\centering
\resizebox{\hsize}{!}{\includegraphics[angle=-90,clip=true]{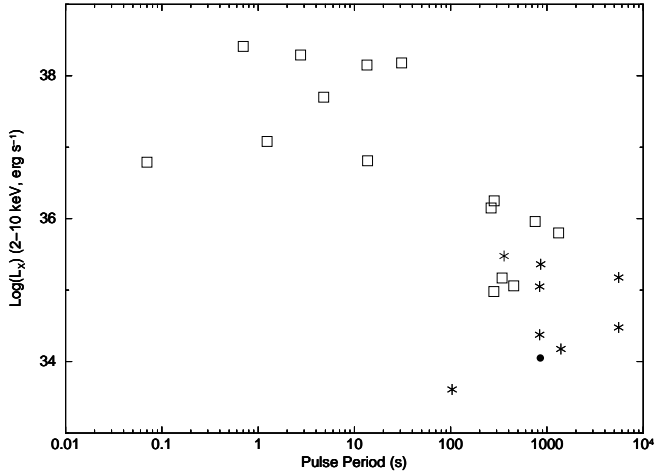}}
\caption{X--ray luminosity (in the 2--10 keV energy range) of the XBPs with a detected thermal excess as a function of the pulse period. The \textit{filled circle} refers to the \XMM\ observation of \RX, \textit{asterisks} refer to other detections of \textit{hot BB} pulsars, \textit{open squares} to all the other bright sources.}
\label{luminosity_period}
\end{figure}

An X--ray excess has been observed also in several other XBPs \citep[see][ for a review]{LaPalombaraMereghetti2006} but, contrary to the previous sources, in their case the fit of this excess with a thermal emission model provided low temperatures ($kT <$ 0.5 keV) and large emitting regions (\textit{R} $>$ 100 km); for this reason, this feature is usually described as a \textit{soft} excess. In Fig.~\ref{luminosity_period} we report the luminosity and pulse period of the XBPs with a detected thermal excess. They are divided in two well distinct groups: the sources in the first group are characterized by high luminosity ($L_{\rm X}\ge10^{37}$ erg s$^{-1}$) and short pulse period (\textit{P} $<$ 100 s), and in most cases they are in close binary systems with an accretion disk; those in the second group have low luminosities ($L_{\rm X}\le10^{36}$ erg s$^{-1}$) and long pulse periods (\textit{P} $>$ 100 s), since they have wide orbits and are wind fed systems. Among the sources of the second group of XBPs, the six pulsars discussed above (reported as \textit{asterisks} and \textit{filled circle}) are the ones that have, at the same time, the lowest luminosities and the longest periods with the only exception of SAX J2103.5+4545, which has the highest luminosity of this group of sources. They are characterized by a \textit{BB} component with a high temperature and a small emission radius, while the others show a \textit{soft} thermal excess with low temperature ($kT <$ 0.5 keV) and large emission area (\textit{R} $\sim$ a few hundred km). The \textit{hot BB} spectral component separates these low--luminosity and long--period sources from all the other pulsars, strongly suggesting that they form a distinct and well defined class of binary pulsars.

Based on our results and on the emission models proposed by \citet{Hickox+04}, the luminosity of \RX\ is too small for an interpretation of the soft excess in terms of reprocessing of the hard X--ray photons in optically thick accreting material. Moreover, both the thermal emission and the hard X--ray reprocessing in a diffuse, optically thin gas around the neutron star are unlikely, since these processes would not give a blackbody spectrum. We therefore favor the interpretation of the soft excess in \RX\ as thermal emission from the neutron star polar cap. If we assume that the source is in the `accretor' status, with matter accretion on the NS surface, the blackbody emitting radius of $\sim$ 130 m is consistent with the expected size of the polar cap. In fact, assuming $M_{\rm NS}$ = 1.4 $M_{\odot}$ and $R_{\rm NS}$ = $10^6$ cm, the source luminosity $L_{\rm X} = 1.2 \times 10^{34}$ erg s$^{-1}$ implies an accretion rate $\dot M \simeq 6.4 \times 10^{13}$ g s$^{-1}$ and, adopting $B_{\rm NS} = 10^{12}$ G, a magnetospheric radius $R_m \simeq 2.5 \times 10^9$ cm \citep{Campana+98}. In this case, based on the relation $R_{col} \sim R_{\rm NS}$ ($R_{\rm NS}/R_m$)$^{0.5}$ \citep{Hickox+04}, we would obtain $R_{col} \sim$ 200 m, comparable to the estimated blackbody emitting radius

The above results suggest that the observed thermal component can be attributed to the emission from the NS polar caps, as already proposed for 4U 0352+309, RX J0146.9+6121 and \TreA. If this description is correct, we would expect to observe some variability of the thermal component along the pulse phase. This hypothesis cannot be confirmed by the present data, since the phase--resolved spectroscopy shows that the significant spectral variability detected along the pulse phase can be due to the \textit{PL} component, with no variation of the \textit{BB} one.

\section{Conclusions}

We have reported the analysis of the data collected by \XMM~in a $\sim$ 28 ks observation of the Be/NS X--ray pulsar \RX. The unabsorbed flux corresponds to a source luminosity $L_{\rm X}\simeq 1.2\times10^{34}$ erg s$^{-1}$ in the 2--10 keV energy range, which is about one order of magnitude lower than the average level of the previous observations. The source pulse period is $P$ = 853.4 $\pm$ 0.2 s, which, together with previous measurements, implies an average pulsar spin--up $\dot P \simeq -2\times10^{-8}$ s s$^{-1}$ in the latest decade.

We could perform the first accurate timing and spectral analysis below 2 keV for this source. The pulse profile shows a complex structure, which is not sinusoidal but does not depend on energy. In the source spectrum we detected a count excess above the main power--law component, well described by a blackbody with $kT_{\rm BB} \simeq$ 1.3 keV and $R_{\rm BB} \simeq$ 130 m, which contributes for $\sim$ 40 \% to the total source luminosity. We found no evidence of a narrow iron K$_{\alpha}$ line between 6 and 7 keV, with an upper limit of 0.2 keV on its equivalent width.

The new \XMM\ detection of the long--period Be binary pulsar \RX\ at a low luminosity level confirms that it belongs to the class of the persistent Be binary pulsars, whose main representatives  are 4U 0352+309 and RX J0146.9+6121. Moreover, there is a full agreement among
the results of the spectral analysis of these three sources; therefore it is reinforced the hypothesis that a thermal excess of blackbody type, with high temperature ($kT >$ 1 keV) and small emission area (\textit{R} $<$ 0.5 km), is a common feature of this type of sources. However, this type of excess is not exclusive of the persistent low--luminosity Be binary pulsars, since also the transient source \TreA\ shows this component. This means that, on one hand, all the low--luminosity Be X--ray binary pulsars show this component but, on the other hand, the presence of such component in a Be binary does not imply that the system is a persistent low--luminosity source.

As in the other two sources, also for \RX\ the blackbody radius is comparable to the estimated size of the NS polar--cap, which suggest that the origin of this component is on the NS surface. However, the phase--resolved spectroscopy does not confirm this scenario, since the observed spectral variability along the pulse period can be attributed to the power--law component and is consistent with a constant thermal component.

\begin{acknowledgements}
This work is based on observations obtained with \XMM, an ESA science mission with instruments and contributions directly funded by ESA Member States  and NASA. The \XMM~data analysis is supported by the Italian Space Agency (ASI/INAF contract I/088/06/0). P.E. thanks the Osio Sotto City Council for support with a ``G. Petrocchi'' fellowship.
\end{acknowledgements}

\bibliographystyle{aa}
\bibliography{biblio}

\end{document}